# Data Descriptor: Nanomaterial datasets to advance tomography in scanning transmission electron microscopy

Barnaby D.A. Levin[1], Elliot Padgett[1], Chien-Chun Chen[2,3], M.C. Scott[2,4], Rui Xu[2], Wolfgang Theis[5], Yi Jiang[6], Yongsoo Yang[2], Colin Ophus[4], Haitao Zhang[7], Don-Hyung Ha[7], Deli Wang[8,9], Yingchao Yu[8], Hector D. Abruña[8], Richard D. Robinson[7], Peter Ercius[4], Lena F. Kourkoutis[1,10], Jianwei Miao[2], David A. Muller[1,10] & Robert Hovden[1]

Electron tomography in materials science has flourished with the demand to characterize nanoscale materials in three dimensions (3D). Access to experimental data is vital for developing and validating reconstruction methods that improve resolution and reduce radiation dose requirements. This work presents five high-quality scanning transmission electron microscope (STEM) tomography datasets in order to address the critical need for open access data in this field. The datasets represent the current limits of experimental technique, are of high quality, and contain materials with structural complexity. Included are tomographic series of a hyperbranched $Co_2P$ nanocrystal, platinum nanoparticles on a carbon nanofibre imaged over the complete 180° tilt range, a platinum nanoparticle and a tungsten needle both imaged at atomic resolution by equal slope tomography, and a through-focal tilt series of PtCu nanoparticles. A volumetric reconstruction from every dataset is provided for comparison and development of post-processing and visualization techniques. Researchers interested in creating novel data processing and reconstruction algorithms will now have access to state of the art experimental test data.

| | |
|---|---|
| **Design Type(s)** | reference design • nanomaterial structure generation objective |
| **Measurement Type(s)** | 3D structure determination assay |
| **Technology Type(s)** | electron tomography |
| **Factor Type(s)** | |
| **Sample Characteristic(s)** | |

[1]School of Applied and Engineering Physics, Cornell University, Ithaca, New York 14853, USA. [2]Department of Physics & Astronomy, and California NanoSystems Institute, University of California, Los Angeles, California 90095, USA. [3]Department of Physics, National Sun Yat-Sen University, Kaohsiung 80424, Taiwan. [4]National Center for Electron Microscopy, Molecular Foundry, Lawrence Berkeley National Laboratory, Berkeley, California 94720, USA. [5]Nanoscale Physics Research Laboratory, School of Physics and Astronomy, University of Birmingham, Edgbaston, Birmingham B15 2TT, UK. [6]Department of Physics, Cornell University, Ithaca, New York 14853, USA. [7]Department of Materials Science and Engineering, Cornell University, Ithaca, New York 14853, USA. [8]Department of Chemistry and Chemical Biology, Cornell University, Ithaca, New York 14853, USA. [9]School of Chemistry and Chemical Engineering, Huazhong University of Science and Technology, Wuhan 430074, China. [10]Kavli Institute for Nanoscale Science, Cornell University, Ithaca, New York 14853, USA. Correspondence and requests for materials should be addressed to R.H. (email: rmh244@cornell.edu).

## Background & Summary

Electron tomography attempts to reconstruct 3D objects from 2D projection images taken at different viewing angles, or tilts—producing the entire internal structure of a specimen or region of interest. Since the first 3D reconstruction from electron micrographs[1], tomography with the scanning transmission electron microscope (STEM) has been widely applied to nanoscale materials[2–12]. Utilizing the sub-angstrom 2D resolution of modern STEM, 3D reconstructions with sub-nanometre and even atomic detail have been demonstrated[13–16]. In the design of advanced nanomaterials, 3D characterization of nano-scale structure offers valuable insight into a material's macroscale function. As a result, demand for nanoscale STEM tomography is high[17].

Routine tomographic methods face several challenges that reduce final quality. The geometry of most specimens and specimen holders restricts tilt range to less than roughly 140°. Commonly referred to as 'the missing wedge', an incomplete tilt-range limits the information available for reconstruction and manifests as an elongation in the final reconstruction[18,19]. Contamination and specimen radiation sensitivity limit the number of viewing angles and the signal to noise, which in turn, restricts the final resolution in 3D[20]. Depth-of-field limits the maximum allowable size of the object that can be reconstructed; a particular problem for aberration corrected STEM[13,21,22].

Recently, efforts towards new reconstruction methods promise higher resolution reconstructions using fewer viewing angles and lower radiation doses than traditional reconstruction algorithms like Weighted Back Projection (WBP) and Simultaneous Iterative Reconstruction Technique (SIRT). New approaches such as the iterative Fourier-based equal slope tomography[23] and compressed sensing inspired algorithms[24,25] have demonstrated success in STEM tomography by improving reconstruction quality with reduced sampling. However, we still lack a fundamental understanding of when and how these algorithms fail. Adopting new algorithms into routine tomography requires thorough investigation[17].

A lack of high-quality, open access data is impeding development and validation of new algorithms and software for 3D reconstruction, visualization, and analysis. Currently, the best tomographic datasets are harboured by a privileged few. Researchers best suited for creating novel data processing and analysis techniques do not readily have access to experimental data.

To address this deficiency, we present five datasets that have pushed the limits of electron tomography. Each dataset was acquired using a unique experimental technique, is of high quality, and contains materials with structural complexity:

Tom_1) Tomography of Hyperbranched $Co_2P$ Nanoparticle: a 150° tomographic tilt series, taken at 2° increments. The $Co_2P$ nanocrystal has a complex morphology of bundled branches that resembles a six-pointed star[26]. This dataset represents a tilt range and increment typical of nano-scale STEM tomography.

Tom_2) 180° Tomography of NPs on Nanofibre: a tilt series taken at 1° tilt increments over the full 180° tilt range of platinum nanoparticles on a graphitized carbon nanofibre support. This dataset provides a complete range of tilts, allowing researchers to better understand the effects of missing information. 16 fast acquisition images were acquired at each tilt, and the experimental signal-to-noise level can be adjusted by averaging different numbers of these images.

Tom_3) Atomic Resolution Tomography of Pt NP: a 145° equal slope tomography tilt series of a single platinum nanoparticle, acquired at atomic resolution, enabling reconstruction of atomic features[15].

Tom_4) Atomic Resolution Tomography of Tungsten Needle: An equal slope tomography tilt series of the tip of a tungsten needle, acquired over the full 180° range, enabling atomic resolution reconstruction[16].

Tom_5) Through-Focal Tomography of Pt-Cu Catalyst: a 138° through-focal tomographic tilt series acquired in an aberration-corrected microscope of Pt-Cu fuel cell catalyst nanoparticles with a complex internal pore structure on an extended carbon support at 3° increments[13]. This dataset overcomes the limited depth of field that accompanies high-resolution aberration corrected imaging[21] by combining through-focal sectioning and tilt-series tomography to reconstruct extended objects.

The datasets include raw tilt series aligned for reconstruction and 3D reconstructions of each specimen—all in an easily readable TIF format.

Combined, these datasets provide a standard, open set of test data for the growing field of tomographic reconstruction and visualization. The datasets allow researchers to rigorously test their algorithms from alignment to reconstruction on real experimental data. The datasets will also find a use as a training tool for scientists new to tomography, a validation tool for 3D tomographic visualization, and a template to seed a future open library of tomographic data.

## Methods

In ADF-STEM electron tomography, a focused electron beam with sub-nanometre diameter is rastered across a sample of interest. Electrons scattered from the sample are recorded using an annular dark field detector, which generates a 2D projection image of the sample[27,28].

The viewing angle is changed by rotating the specimen and a series of projection images from different angles is acquired. For the vast majority of electron microscopes, a single axis of rotation is permitted by the stage, although more complex tilt geometries have been demonstrated[17,29,30]. After each successive tilt during the experiment, the specimen moves relative to the electron beam and must be re-centred. The specimen can only be re-centred approximately at the time of acquisition and so the data is said to be 'misaligned'. The end result of a tomographic experiment is a set of 'misaligned' STEM images corresponding to a specific specimen tilt.

Aligning the STEM images prior to reconstruction is vital to establish a common axis of rotation (i.e., tilt-axis) in the image series. Alignment methods include the use of fiducial markers[31,32], cross correlation[33], and centre of mass. Once aligned, a reconstruction algorithm is used to generate a 3D reconstruction of the sample.

The basic steps of the tomographic method are illustrated in Fig. 1.

### Tom_1: Tomography of hyperbranched $Co_2P$ nanoparticle

**Sample preparation.** Hyperbranched $Co_2P$ nanocrystal synthesis methods and scientific relevance are discussed in detail by Zhang et al.[26] Samples were prepared for tomographic analysis by pipetting a drop of organic solution containing $Co_2P$ nanocrystals onto the surface of a copper TEM grid coated with an amorphous carbon film. Once the organic solution had dried, $Co_2P$ nanocrystals were dispersed over the grid. A drop of a solution of gold nanoparticles in water was then pipetted onto the grid, and allowed to dry. The gold nanoparticles were used as fiducial markers to align the tilt series.

**Data acquisition.** The tomographic tilt series of $Co_2P$ nanocrystals was acquired using an FEI Tecnai F20 scanning transmission electron microscope (STEM) at Cornell University. The microscope was operated at an accelerating voltage of 200 kV, with a convergence semi-angle of 9.6 mrad, and beam current of ~8–10 pA. This yields a nominal 2D resolution of up to 1.6 Å for STEM annular dark field (ADF) images. The tomographic tilt series was acquired over a 150° range at 2° intervals using a high angle annular dark-field (HAADF) detector. The scale in each image is ~0.71 nm per pixel.

**Alignment and reconstruction.** Each of the 76 projections in the tilt series, tiltser_Co2P.tif, has been aligned to a fiducial particle close to the $Co_2P$ nanocrystal using manual alignment techniques (except projections 50 and 51, which are blank to correct for 4° goniometer backlash during acquisition). Similarly, the tilt axis was determined by manually choosing the axis of rotation that minimized artifacts and maximized detail in the final reconstruction. We provide an example reconstruction, recon_Co2P.tif, produced from tiltser_Co2P.tif using the SIRT algorithm.

### Tom_2: 180° Tomography of nanoparticles on nanofibre

**Sample preparation.** Graphitized nanofibres, loaded with platinum nanoparticles at 10 wt. % were dispersed in a methanol solution and dried onto the tip of a tungsten omniprobe needle. The needle was inserted into a Fischione 2050 On-Axis Rotation Tomography Holder for data acquisition.

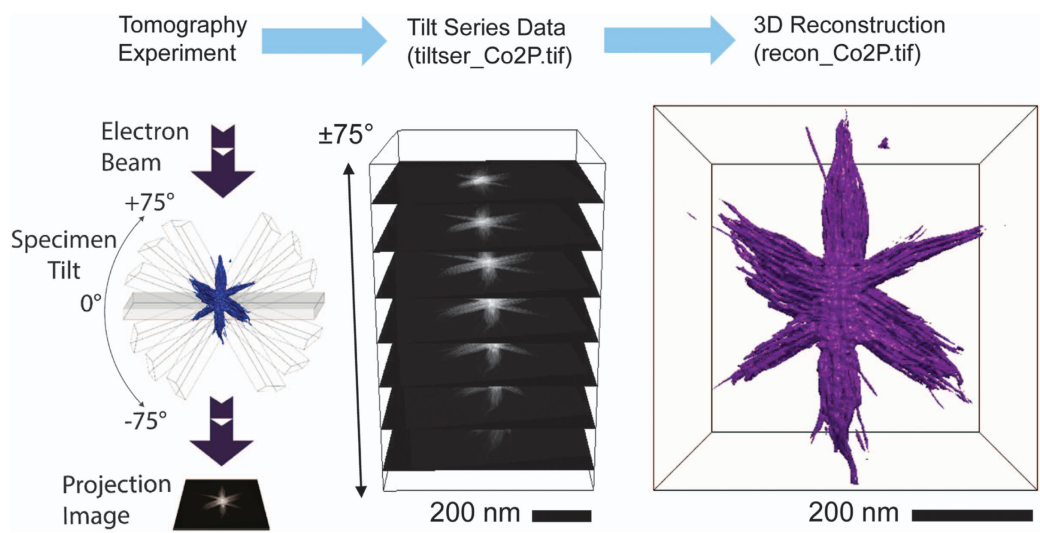

**Figure 1.** Illustration of electron tomography data acquisition and reconstruction process. Series of 2D images acquired of object of unknown 3D structure at different viewing angles. Images shown are from tiltser_Co2P.tif. 2D images combined into image stack ordered by viewing angle i.e., a tilt series. Tilt series is aligned, and reconstruction algorithm is applied to produce 3D reconstruction of object. A 3D isosurface visualization of recon_Co2P.tif is shown as an example rendered using tomviz.

**Data acquisition.** One 93° tilt series and one 95° tilt series were acquired with an offset of ~85.3° between the viewing angle of the first image of the first series, and the viewing angle of the first image of the second series. The two tilt series together therefore cover slightly more than the full 180° tilt range. The overlapping region between the two tilt series was used to align them together in post-processing. The angular increment for each tilt series was 1°. In order to reduce scan noise and thus improve signal to noise ratio, 16 images, each with a 1 μs per pixel dwell time, were recorded at each viewing angle and saved as an image stack to be aligned in post processing. Data was acquired using an FEI Tecnai F20 scanning transmission electron microscope (STEM) at Cornell University. The microscope was operated in low angle annular dark field (LAADF) mode an accelerating voltage of 200 kV with a probe current of approximately 5 pA. A convergence angle of ~6.9 mrad was used to optimize resolution over a large depth of field. Images were acquired with 1024 × 1024 pixels. Non-orthogonality in probe scan direction was observed and was corrected in all images by applying a 0.6 degree shear parallel to the tilt axis with linear interpolation. The field of view in each image was 363.52 nm.

**Alignment and reconstruction.** Each of the 1 μs per pixel image stacks of 16 images was aligned by cross correlation. Each stack was then summed to form a single image. The single images from both tilt series were then combined into a single 180° image stack, tiltser_180.tif, for reconstruction, with duplicate viewing angles discarded. The tilt series was aligned using a centre of mass method. A 3D reconstruction of the data, recon_180.tif, was produced using a weighted back projection algorithm. Data illustrated in Fig. 2a.

**Tom_3: Atomic resolution tomography of platinum nanoparticle**
**Sample preparation.** Platinum nanoparticles were deposited onto a grid consisting of a 5-nm-thick silicon nitride membrane with dimensions of 100 μm × 1500 μm, supported on a 100 μm-thick silicon frame designed for loading into a TEM (TEMwindows.com). High temperature coating of a 1–2 nm carbon layer was applied to mitigate charging effects due to the electron beam. The grid was then loaded onto a Fischione 2020 tomographic sample holder for data acquisition in the TEM.

**Data acquisition.** The tomographic tilt series of platinum nanoparticles was acquired using an uncorrected FEI Titan STEM at the University of California, Los Angeles. The microscope was operated with a beam energy of 200 keV, a 100 pA probe current, and a 10.7 mrad convergence semi-angle. A tilt series of 104 projections was acquired from a platinum nanoparticle with equal-slope increments and a tilt range of ±72.6°.

**Alignment and reconstruction.** The images in the tilt series, tiltser_PtNP.tif, were aligned using a centre of mass (CM) alignment method after background subtraction and removal[15]. We present a reconstruction of this data, recon_PtNP.tif, produced using the equal slope tomography (EST) iterative algorithm, a method described by Miao *et al.*[23] No Fourier filters were applied to the final reconstruction. Data illustrated in Fig. 2b.

**Tom_4: Atomic resolution tomography of tungsten needle**
**Sample preparation.** A 99.95% pure tungsten wire was annealed under tension, creating a large crystalline domain with the [011] crystallographic axis aligned along the wire axis. The wire was electrochemically etched in a NaOH solution to form a sharp tip with a < 10 nm diameter. The wire was plasma cleaned in an $Ar/O_2$ gas mixture and then heated to 1,000 °C under vacuum (~10 − 5 Pa) to remove the oxide layer generated by the plasma cleaning. The wire was mounted in a 1 mm sample puck compatible with the TEAM microscope stage.

**Data acquisition.** Tomographic data was acquired using the TEAM I at the National Center for Electron Microscopy. The microscope was operated at 300 kV beam voltage in ADF-STEM mode with a convergence semi-angle of ~ 30 mrad and a ~ 70 pA beam current. The tomography rotation axis was aligned to the wire axis [011]. An equally sloped tomographic tilt series of 62 images, covering the complete angular range of ±90° was acquired from the tungsten needle sample. Two images of 1024 × 1024 pixels each with 6 μs per pixel dwell time and 0.405 Å pixel resolution were acquired at each angle in order to correct for drift. The TEAM stage, which is a tilt-rotate design with full 360° rotation about both axes, enabled rotation around the [011] crystalline axis.

**Alignment and reconstruction.** Raw experimental data can be found in tiltser_W.zip, which contains tif stacks of the two images acquired at each viewing angle, as described above. In addition, we provide an aligned tilt series, tiltser_W.tif. In the raw data, the tilt axis has a different in-plane orientation at each viewing angle, and in order to obtain an aligned tilt series, this was corrected by using Fourier methods to align the tilt direction along the image horizontal in every image. Both sample drift and scan distortion were corrected for in all images in the tilt series using Fourier techniques[34]. The tilt series was then aligned using a centre of mass method, with a mask applied to remove background noise. The tilt series was cropped in order to only feature the tip of the needle, which remained within the depth of focus throughout data acquisition. We present a reconstruction of the data, recon_W.tif, produced from

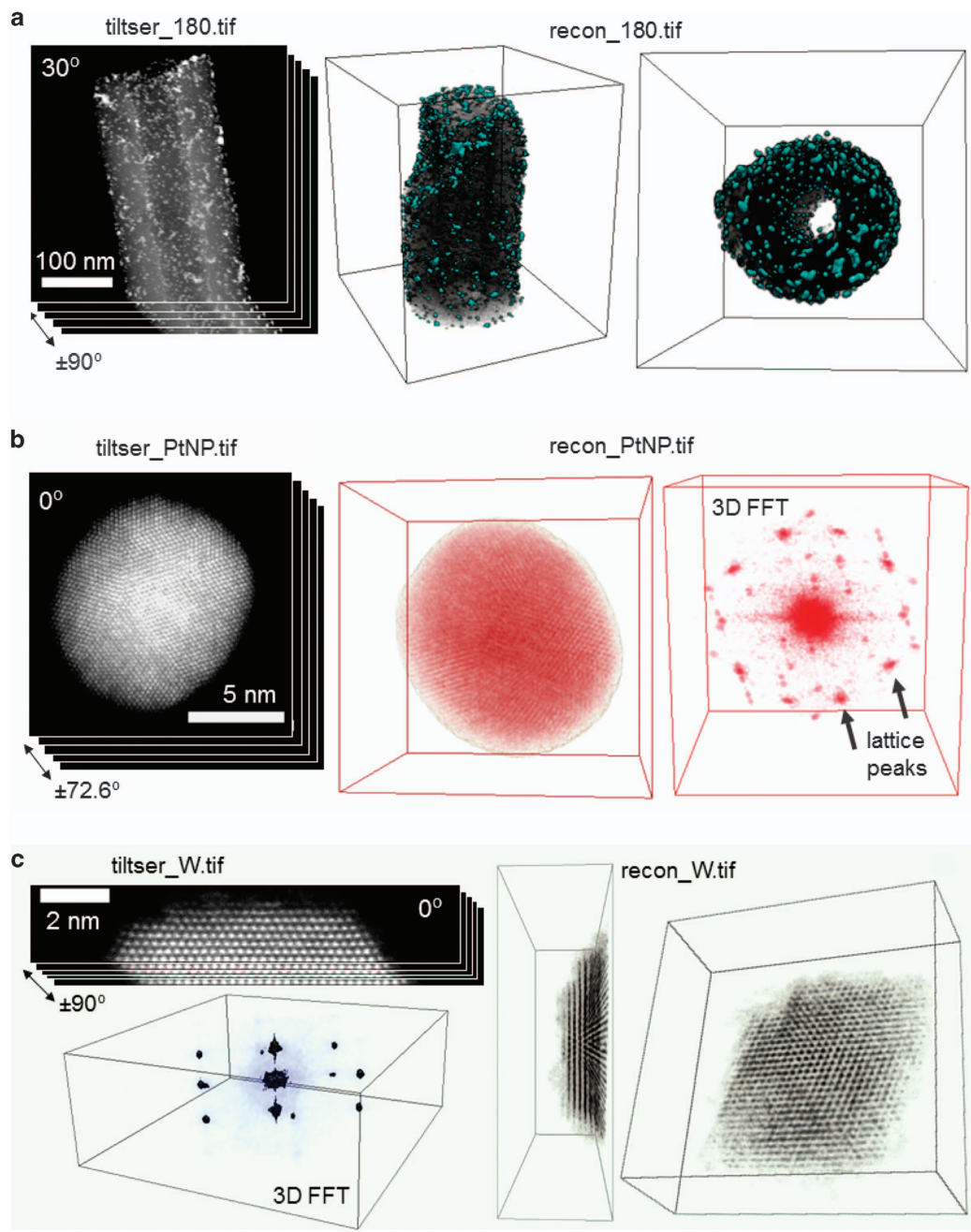

**Figure 2.** Illustrations of tilt series and sample reconstructions. (**a**) Sample image from tiltser_180.tif. Mixed 3D volume/isosurface visualizations of recon_180.tif show exterior of fibre, with nanoparticles visible on exterior, and hollow interior of nanofibre, containing nanoparticles. (**b**) Sample image from tiltser_PtNP.tif. Mixed 3D volume/isosurface visualization of recon_PtNP.tif and volume visualization of 3D Fourier transform of recon_PtNP.tif, showing platinum reciprocal lattice spots. (**c**) Sample image from tiltser_W.tif. Mixed 3D volume and isosurface visualization of recon_W.tif and of the 3D Fourier transform (cropped) of recon_W.tif, showing tungsten reciprocal lattice spots. All 3D visualizations produced using tomviz.

tiltser_W.tif using the equal slope tomography iterative algorithm. The alignment and reconstruction process is explained in detail by Xu *et al.*[16] Data illustrated in Fig. 2c.

### Tom_5: Through-focal tomography of Pt-Cu catalyst
**Sample preparation.** The through focal tilt series was acquired on PtCu nanoparticles on a 3D Vulcan carbon support. The synthesis methods and scientific relevance of the nanoparticles as a fuel-cell electrocatalyst are discussed in detail by Wang *et al.*[35] To prepare for observation in the electron microscope, the particles were suspended in ethanol and pipetted onto a copper TEM grid with an ultra-thin, holey carbon support film.

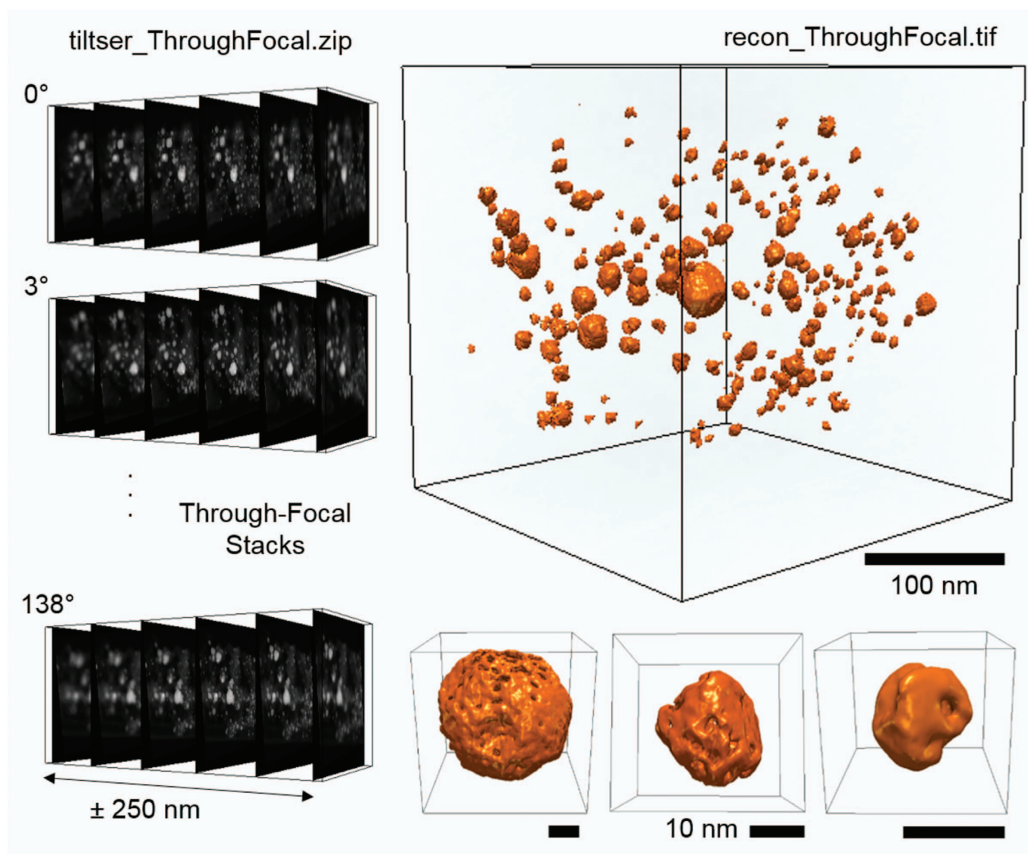

**Figure 3.** Illustration of raw data and sample reconstruction for Tom_5. A through-focal image series must be acquired at each viewing angle in through-focal tomography. Files 018.tif, 072.tif, and 120.tif are shown as examples. Through-focal tomography allows objects from an extended field of view to be reconstructed at high resolution in an aberration corrected STEM. A 3D isosurface visualization of the full view of PtCu nanoparticles on an extended carbon support in recon_ThroughFocal.tif is shown, along with high resolution 3D visualizations of individual PtCu particles in the reconstruction. All visualizations produced using tomviz.

**Data acquisition.** The through-focal tomographic tilt series of de-alloyed PtCu nanoparticles on an extended 3D carbon support was acquired using TEAM I at the National Center for Electron Microscopy; a tool that provides attributes to best demonstrate the advantages of this technique. Its large convergence angle provides high lateral resolution (<0.78 Å) and a small depth-of-field (~6 nm) at 300 kV accelerating voltage. Shadowing from the TEM grid limited tilts from −68° to +71° along our chosen axis of rotation.

The tomographic data was acquired over a 138° tilt range using a high angle annular dark field (HAADF) detector. The 30 mrad convergence angle provided a continuum of information in the through-focal CTF that spanned a ±1.72° wedge at low and medium frequencies. A 3° tilt increment was chosen to match the convergence angle. The PtCu nanoparticles decorate a 3D Vulcan carbon support with an extended structure that far exceeds the microscope's depth of field—making it impossible to image multiple particles in-focus within a single field of view. At every tilt a 26 image through-focal series was taken over ±250 nm defocus with 20 nm focal steps in order to ensure all objects were imaged in focus. The microscope defocus steps are calibrated from a through-focal stack (Fig. 3). Each image had a 0.38 nm per pixel lateral resolution.

**Alignment and reconstruction.** A five-dimensional alignment of the raw data in tiltser_ThroughFocal.zip was required: transverse x-y alignment, focal z-alignment, tilt axis rotation and shift. A fiducial particle was used to align each through-focal stack in their respective x-y direction. The focal z-alignment for each focal stack was determined by identification of the best focus image to a fiducial particle. Within each focal stack a cross-correlation alignment was used to reduce the small amounts of drift during the acquisition. After alignment, the data was reweighted in Fourier space by dividing with the microscope's contrast transfer function (CTF) approximated by a 300 keV 30 mrad aberration-free probe plus a Wiener constant of 5 times the max CTF value. After this light deconvolution, each through-focal stack was mapped onto a universal Fourier space by bilinear extrapolation. This extrapolation distributes the complex value of an input point to its four nearest neighbors on the output Cartesian grid with a weighted average of points from all the through-focal stacks. A direct inverse 3D Fourier transform provided the final reconstruction, recon_ThroughFocal.tif. This method is described by Hovden *et al.*[13]. It should be noted that the alignment of each through-focal stack generated excess blank images for reference. Thus in the raw data provided, there are 43 images per stack; 26 images of the PtCu nanoparticles, and 17 blank reference images.

### Code availability

Code equivalent to that used to reconstruct the data in Tom_1 and Tom_2 is available as part of the open source Tomviz software package at www.tomviz.org. Code used to reconstruct the data in Tom_3 is freely available online at http://www.physics.ucla.edu/research/imaging/EST/. Code used to reconstruct the data in Tom_4 is freely available online at http://www.physics.ucla.edu/research/imaging/3Datoms.

Code used to reconstruct the data in Tom_5 is available in the Supplementary Information to this paper (Supplementary File 1). Alignment tools are available as part of the open source Tomviz software package at www.tomviz.org, and the open source IMOD software package at http://bio3d.colorado.edu/imod/.

### Data Records

The datasets described in this paper are available at Figshare (Data Citation 1). All files are provided in 16-bit tif image format. Table 1 describes the content of the raw tilt series datasets. Table 2 describes tilt series derived from raw data, from which reconstructions are in turn derived, and Table 3 describes the sample reconstructions derived from the raw datasets that we have provided. All tilt series have their axis of rotation along the x-axis (horizontal) of the images.

| Data ID | Raw Dataset | Tomography Method | No. Images (Stacks) | Image Size (Pixels) | Angle Step | Angle Range | Pixel Size |
|---|---|---|---|---|---|---|---|
| Tom_1 | tiltser_Co2P.tif | Traditional | 76 (1) | 1157 × 1157 | 2° | 146° | 0.71 nm |
| Tom_2 | tiltser_180.zip | 180° | 3040 (190) | 947 × 1033 | 1° | I) 93° II) 108° | 0.36 nm |
| Tom_3 | tiltser_PtNP.tif | EST | 109 (1) | 401 × 401 | N/A | 145° | 0.035 nm |
| Tom_4 | tiltser_W.zip | EST | 126 (63) | 1024 × 1024 | N/A | 180° | 0.0405 nm |
| Tom_5 | tiltser_ThroughFocal.zip | Through-Focal | 2021 (47) | 1145 × 1145 | 3° | 138° | 0.38 nm |

**Table 1.** Raw tilt series metadata.

| Data ID | Derived Dataset (Tilt Series) | Data Processing Methods | No. Images | Image Size (Pixels) | Pixel Size |
|---|---|---|---|---|---|
| Tom_2 | tiltser_180.tif | Sum and combine image stacks. | 180 | 947 × 1033 | 0.36 nm |
| Tom_4 | tiltser_W.tif | Align images. Correct drift and distortions. Sum and combine image stacks. Remove noise. Crop tilt series. | 62 | 54 × 331 | 0.0405 nm |

**Table 2.** Derived tilt series metadata. Tilt increments (angle step) and angular ranges as in Table 1 above.

| Data ID | Derived Dataset (Reconstruction) | Reconstruction Method | Reconstruction Size (Voxels) | Voxel Size |
|---|---|---|---|---|
| Tom_1 | recon_Co2P.tif | SIRT | 579 × 579 × 579 | 1.42 nm |
| Tom_2 | recon_180.tif | WBP | 517 × 517 × 522 | 0.72 nm |
| Tom_3 | recon_PtNP.tif | EST Iterative Algorithm | 241 × 241 × 241 | 0.058 nm |
| Tom_4 | recon_W.tif | EST Iterative Algorithm | 255 × 255 × 105 | 0.053 nm |
| Tom_5 | recon_ThroughFocal.tif | Direct Fourier Transform | 1025 × 1025 × 1025 | 0.38 nm |

**Table 3.** Derived reconstruction metadata.

**Further information on Tom_2: 180 degree tomography of NPs on nanofibre**

tiltser_180.zip contains 190.tif stacks. Each stack is a series of sixteen images of one viewing angle, with each image acquired at 1 µs per pixel dwell time. The images in the each of stacks have been aligned by cross correlation. The image stacks have also been aligned with each other, allowing users to construct their own aligned tilt series from the image stacks. A full file listing for tiltser_180.zip is given below. Roman numerals indicate a stack from the first or the second tilt series. The number in the label indicates the viewing angle given by the microscope goniometer.

| | | | | |
|---|---|---|---|---|
| I_00.tif | I_12.tif | I_-23.tif | I_35.tif | I_-46.tif |
| I_01.tif | I_-12.tif | I_24.tif | I_-35.tif | I_47.tif |
| I_-01.tif | I_13.tif | I_-24.tif | I_36.tif | II_00.tif |
| I_02.tif | I_-13.tif | I_25.tif | I_-36.tif | II_01.tif |
| I_-02.tif | I_14.tif | I_-25.tif | I_37.tif | II_-01.tif |
| I_03.tif | I_-14.tif | I_26.tif | I_-37.tif | II_02.tif |
| I_-03.tif | I_15.tif | I_-26.tif | I_38.tif | II_-02.tif |
| I_04.tif | I_-15.tif | I_27.tif | I_-38.tif | II_03.tif |
| I_-04.tif | I_16.tif | I_-27.tif | I_39.tif | II_-03.tif |
| I_05.tif | I_-16.tif | I_28.tif | I_-39.tif | II_04.tif |
| I_-05.tif | I_17.tif | I_-28.tif | I_40.tif | II_-04.tif |
| I_06.tif | I_-17.tif | I_29.tif | I_-40.tif | II_05.tif |
| I_-06.tif | I_18.tif | I_-29.tif | I_41.tif | II_-05.tif |
| I_07.tif | I_-18.tif | I_30.tif | I_-41.tif | II_06.tif |
| I_-07.tif | I_19.tif | I_-30.tif | I_42.tif | II_-06.tif |
| I_08.tif | I_-19.tif | I_31.tif | I_-42.tif | II_07.tif |
| I_-08.tif | I_20.tif | I_-31.tif | I_43.tif | II_-07.tif |
| I_09.tif | I_-20.tif | I_32.tif | I_-43.tif | II_08.tif |
| I_-09.tif | I_21.tif | I_-32.tif | I_44.tif | II_-08.tif |
| I_10.tif | I_-21.tif | I_33.tif | I_-44.tif | II_09.tif |
| I_-10.tif | I_22.tif | I_-33.tif | I_45.tif | II_-09.tif |
| I_11.tif | I_-22.tif | I_34.tif | I_-45.tif | II_10.tif |
| I_-11.tif | I_23.tif | I_-34.tif | I_46.tif | II_-10.tif |
| II_11.tif | II_-18.tif | II_26.tif | II_-33.tif | II_41.tif |
| II_-11.tif | II_19.tif | II_-26.tif | II_34.tif | II_-41.tif |
| II_12.tif | II_-19.tif | II_27.tif | II_-34.tif | II_42.tif |
| II_-12.tif | II_20.tif | II_-27.tif | II_35.tif | II_-42.tif |
| II_13.tif | II_-20.tif | II_28.tif | II_-35.tif | II_43.tif |
| II_-13.tif | II_21.tif | II_-28.tif | II_36.tif | II_-43.tif |
| II_14.tif | II_-21.tif | II_29.tif | II_-36.tif | II_44.tif |
| II_-14.tif | II_22.tif | II_-29.tif | II_37.tif | II_-44.tif |
| II_15.tif | II_-22.tif | II_30.tif | II_-37.tif | II_45.tif |
| II_-15.tif | II_23.tif | II_-30.tif | II_38.tif | II_-45.tif |
| II_16.tif | II_-23.tif | II_31.tif | II_-38.tif | II_46.tif |
| II_-16.tif | II_24.tif | II_-31.tif | II_39.tif | II_-46.tif |
| II_17.tif | II_-24.tif | II_32.tif | II_-39.tif | II_47.tif |
| II_-17.tif | II_25.tif | II_-32.tif | II_40.tif | II_-47.tif |
| II_18.tif | II_-25.tif | II_33.tif | II_-40.tif | II_48.tif |

**Further information on Tom_3: Atomic resolution tomography of platinum nanoparticle**

The images in tiltser_PtNP.tif were acquired using the equal slope tomography method. Rather than acquiring images at a fixed angular increments as in traditional tomography, the images are acquired at viewing angles that give equal increments of the slope of the Fourier transformed image planes in Fourier space[23]. The viewing angles (in degrees) associated with each of the 109 images in the tif stack are listed below:

| | | | | |
|---|---|---|---|---|
| 72.646 | 45.000 | 17.354 | − 20.556 | − 46.848 |
| 71.030 | 44.091 | 15.709 | − 22.109 | − 47.816 |
| 69.444 | 43.152 | 14.036 | − 23.629 | − 48.814 |
| 67.891 | 42.184 | 12.339 | − 25.115 | − 49.844 |
| 66.371 | 41.186 | 10.62 | − 26.565 | − 50.906 |
| 64.885 | 40.156 | 8.8807 | − 27.979 | − 52.001 |
| 63.435 | 39.094 | 7.125 | − 29.358 | − 53.130 |
| 62.021 | 37.999 | 5.3558 | − 30.700 | − 54.293 |
| 60.642 | 36.870 | 3.5763 | − 32.005 | − 55.491 |
| 59.300 | 35.707 | 1.7899 | − 33.275 | − 56.725 |
| 57.995 | 34.509 | 0.0000 | − 34.509 | − 57.995 |
| 56.725 | 33.275 | − 1.7899 | − 35.707 | − 59.300 |
| 55.491 | 32.005 | − 3.5763 | − 36.870 | − 60.642 |
| 54.293 | 30.700 | − 5.3558 | − 37.999 | − 62.021 |
| 53.130 | 29.358 | − 7.1250 | − 39.094 | − 63.435 |
| 52.001 | 27.979 | − 8.8807 | − 40.156 | − 64.885 |
| 50.906 | 26.565 | − 10.620 | − 41.186 | − 66.371 |
| 49.844 | 25.115 | − 12.339 | − 42.184 | − 67.891 |
| 48.814 | 23.629 | − 14.036 | − 43.152 | − 69.444 |
| 47.816 | 22.109 | − 15.709 | − 44.091 | − 71.030 |
| 46.848 | 20.556 | −17.354 | −45.000 | −72.646 |
| 45.909 | 18.970 | −18.970 | −45.909 | |

**Further information on Tom_4: Atomic resolution tomography of tungsten needle**

The images in tiltser_W.zip were acquired using the equal slope tomography method. Rather than acquiring images at a fixed angular increments as in traditional tomography, the images are acquired at viewing angles that give equal increments of the slope of the Fourier transformed image planes in Fourier space[23]. The viewing angles (in degrees) associated with each of the 63 image stacks is given in the filename of each image. A list of files in tiltser_W.zip is given below. In order to obtain a reconstruction from the raw images, they must be combined into a tilt series, and aligned (see Tom_4 Atomic Resolution Tomography of Tungsten Needle under Methods).

| | | | | |
|---|---|---|---|---|
| 0.tif | 39.tif | 69.3.tif | 116.4.tif | 147.8.tif |
| 3.5.tif | 41.1.tif | 72.5.tif | 119.2.tif | 150.5.tif |
| 7.1.tif | 43.1.tif | 75.9.tif | 121.9.tif | 153.3.tif |
| 10.6.tif | 44.9.tif | 79.3.tif | 124.4.tif | 156.2.tif |
| 14.tif | 46.8.tif | 82.7.tif | 126.7.tif | 159.1.tif |
| 17.3.tif | 48.8.tif | 86.3.tif | 128.9.tif | 159.3.tif |
| 20.5.tif | 50.8.tif | 89.8.tif | 131.1.tif | 164.3.tif |
| 23.6.tif | 53.1.tif | 93.5.tif | 134.9.tif | 167.8.tif |
| 26.5.tif | 55.4.tif | 97.tif | 136.7.tif | 168.tif |
| 29.3.tif | 57.9.tif | 100.4.tif | 138.7.tif | 175.9.tif |
| 32.tif | 60.6.tif | 103.9.tif | 140.8.tif | 180.tif |
| 34.5.tif | 63.3.tif | 107.2.tif | 143.tif | |
| 36.8.tif | 66.3.tif | 110.4.tif | 145.3.tif | |

**Further information on Tom_5: Through-focal tomography of Pt-Cu catalyst**

tiltser_ThroughFocal.zip contains 47 tif stacks. Each stack is an individual through focal series taken at a different viewing angle. Each of these tif stacks contains 43 images, 26 images of the sample, and 17 blank reference images for through-focal alignment. A reconstruction incorporating all of the information available in the data must be produced directly from the images in each of the 47 image stacks, rather than by combining the stacks into a single file tilt series as for the datasets above.

The title of each tif stack is the viewing angle in degrees that the through focal series was acquired at, measured from the first viewing angle. For example, the first through-focal tif stack, taken at 0°, is named 000.tif. The second tif stack, taken at a tilt of 3° relative to the first is labelled 003.tif. The focal increment between each image in each tif stack is 20 nm.

A full file listing for tiltser_ThroughFocal.zip is given below:

| | | | | |
|---|---|---|---|---|
| 000.tif | 030.tif | 060.tif | 090.tif | 120.tif |
| 003.tif | 033.tif | 063.tif | 093.tif | 123.tif |
| 006.tif | 036.tif | 066.tif | 096.tif | 126.tif |
| 009.tif | 039.tif | 069.tif | 099.tif | 129.tif |
| 012.tif | 042.tif | 072.tif | 102.tif | 132.tif |
| 015.tif | 045.tif | 075.tif | 105.tif | 135.tif |
| 018.tif | 048.tif | 078.tif | 108.tif | 138.tif |
| 021.tif | 051.tif | 081.tif | 111.tif | |
| 024.tif | 054.tif | 084.tif | 114.tif | |
| 027.tif | 057.tif | 087.tif | 117.tif | |

Image 39 in stack 120.tif was not used as part of the direct Fourier reconstruction of the data because it was observed to contain a large scan distortion.

### Technical Validation

The electron microscopes used to acquire the datasets described in this paper were professionally maintained and aligned for optimal imaging conditions prior to dataset acquisition. For Tom_1, Tom_2, Tom_3, and Tom_4 an appropriately sized C2 aperture was selected for data acquisition in order to produce a depth of field that extended over the entire height of the object (or section of the object) to be imaged and reconstructed. For Tom_5 a small depth of field enhanced the through focal technique by providing more 3D information at every tilt. The images in each tilt series presented in this paper have been aligned. The high quality sample reconstructions we provide validate the accuracy and quality of the raw data. No obvious signs of morphological distortions that would indicate poor data acquisition are visible in either the raw data or in the reconstructions themselves. Sinograms of the reconstruction were also inspected for proper alignment to ensure that reconstructions of the highest quality were obtained[18].

### Usage Notes

Collectively, these tomographic datasets of nanoscale materials provide a standard for the development and validation of new 3D imaging methods—from alignment, to reconstruction, to visualization and analysis. Their uses are diverse. The tilt series data can be intentionally degraded by adding misalignment or noise to explore its influence on a particular reconstruction algorithm. In Tom_2 true experimental noise can be added by discarding images within each tilt, thereby reducing the signal to noise ratio. The effects of increasing missing wedge size or tilt increment size can be explored by removing projections from the 180° tilt range in Tom_2 or Tom_4. The high resolution of data in Tom_3 and Tom_4 provide lattice peaks in the 3D Fourier transform of the final reconstruction. These lattice peaks may have appeal to understanding post processing filters. Lastly, Tom_5 provides exploration the limited depth of field that accompanies a new generation of aberration-electron microscopes and its influence on tomography. Each reconstruction provides a playground for visualization and standards for comparison. Tom_1 in particular has an intricate morphology and aesthetic beauty.

This manuscript illustrates the steps necessary to acquire, align, and reconstruct data from nanoscale specimens at the highest quality. The educational utility of the openly available datasets presented here toward training new scientists in electron tomography should not be understated.

The tilt-series data is best viewed using 2D image processing software such as ImageJ, Fiji, or Cornell Spectrum Imager[36]. The reconstructed datasets require 3D visualization software. The open source software tomviz (www.tomviz.org) was used to produce the data visualizations included in this paper[37]. Alternatives include the free to use UCSF Chimera and commercial tools. Datasets and reconstructions may be viewed on Windows, Mac OSX, and Linux operating systems. We recommend a RAM of at least twice the size of the file being viewed for best performance.

### Data Citation
1. Levin, B. D. A. *et al. Figshare* https://dx.doi.org/10.6084/m9.figshare.c.2185342 (2016).

### Acknowledgements

B.L., D.M., R.H. acknowledge support from DOE Office of Science contract DE-SC0011385. E.P. acknowledges support from NSF GRFP grant number is DGE-1144153, and from General Motors (GM), and acknowledges Zhongyi (Vic) Liu of GM for providing materials. Y.J. acknowledges support from DOE grant number DE-SC0005827. J.M. acknowledges support from DOE Office of Science contract DE-SC0010378 and NSF (DMR-1437263). This work made use of the Cornell Center for Materials Research (CCMR) Facilities supported by the National Science Foundation under Award Number DMR-1120296, and the National Center for Electron Microscopy, Lawrence Berkeley National Laboratory, which is supported by the U.S. Department of Energy under Contract no. DE-AC02-05CH11231. We thank John Grazul for TEM assistance in CCMR.

### Author Contributions
B.L. & R.H. wrote and prepared the manuscript. R.H., B.L. & L.K. contributed to acquiring and reconstructing data for Tom_1. E.P., B.L., D.M. & R.H. contributed to acquiring and reconstructing data for Tom_2. M.S., C.C., & J.M. contributed to acquiring and reconstructing data for Tom_3. R.X., M.S., C.C., W.T., Y. Yang, C.O., P.E. & J.M. contributed to acquiring and reconstructing data for Tom_4. R.H., P.E., Y.J. & D.M. contributed to acquiring and reconstructing data for Tom_5. H.Z., D.H., R.R & D.W, Y.Yu., H.A. synthesized particles. All authors reviewed the manuscript.

### Additional Information
Supplementary information accompanies this paper at http://www.nature.com/sdata

**Competing financial interests:** The authors declare no competing financial interests.

**How to cite this article:** Levin, B. D. A. *et al.* Nanomaterial datasets to advance tomography in scanning transmission electron microscopy. *Sci. Data* 3:160041 doi: 10.1038/sdata.2016.41 (2016).